\documentclass[twocolumn]{mnras}
\usepackage[T1]{fontenc}
\usepackage[latin9]{inputenc}
\usepackage{amssymb}
\usepackage{graphicx}
\usepackage{wasysym}
\usepackage{esint}
\newcommand{\pulsar}{PSR~J1231$-$1411}
\newcommand{\pulsarsec}{PSR~J1023$+$0038}

\makeatletter

\providecommand{\tabularnewline}{\\}

\usepackage[normalem]{ulem}

\makeatother

\begin{document}

\title{X-ray bounds on the r-mode amplitude in millisecond pulsars}

\author[K. Schwenzer et al.]{Kai Schwenzer$^{1}$, Tu\u{g}ba Boztepe$^{2}$, Tolga G\"uver$^{2,3}$, 
and Eda Vurgun$^{2}$
\\
$^{1}$Theoretical Astrophysics (IAAT), Eberhard Karls University
of T\"ubingen, T\"ubingen 72076, Germany \\
$^{2}$Istanbul University, Science Faculty, Department of Astronomy
and Space Sciences, Beyaz\i t, 34119, Istanbul, Turkey\\
$^{3}$ Istanbul University Observatory Research and Application Center, Beyaz\i t, 34119, Istanbul,Turkey}

\maketitle

\begin{abstract}
r-mode astroseismology provides a unique way to study the internal
composition of compact stars. Due to their precise timing, recycled
millisecond radio pulsars present a particularly promising class of
sources. Although their thermal properties are still poorly constrained,
X-ray data is very useful for astroseismology since r-modes could
strongly heat a star. Using known and new upper bounds on the temperatures
and luminosities of several non-accreting millisecond radio pulsars
we derive bounds on the r-mode amplitude as low as $\alpha\lesssim10^{-8}$
and discuss the impact on scenarios for their internal composition.
\end{abstract}

\section{Introduction}

Astroseismology is an ideal method to probe the otherwise inaccessible
interiors of stars. In case of compact stars their oscillations generally
cannot be detected directly via an electromagnetic signal and require the connection to other astrophysical
observables. Such a connection is provided by r-modes \cite{Papaloizou:1978zz,Andersson:1997xt,Friedman:1997uh,Lindblom:1998wf,Andersson:1998ze,Andersson:2000mf}
which are unstable due to the Friedman-Schutz mechanism \cite{Friedman:1978hf}
and emit gravitational waves spinning the star down. An important
property of r-modes is that due to their unstable nature, the saturation
of their amplitude at a finite value also dissipates a significant
amount of energy that heats the star. As discussed in \cite{Alford:2013pma}
this could be particularly important in recycled millisecond radio
or high-energy pulsars, since r-modes with amplitudes that would explain
the known spin-down behavior would heat these otherwise rather cold
sources to large surface temperatures $\gtrsim\!10^{6}\,{\rm K}$.
Temperature measurements or bounds on these sources therefore allow
us to constrain the r-mode amplitude and to distinguish different
scenarios \cite{Alford:2013pma}.

For recycled, non-accreting millisecond pulsars detailed X-ray measurements
are still scarce due to their faintness \cite{Becker:1998yf,Prinz:2015jkd}.
Moreover, the pulsar beaming mechanism, creating a hot spot on the
surface and hard x-ray components from the magnetosphere, complicates
the X-ray analysis. Because of this an actual surface temperature
measurement is only available for the closest source J0437$-$4715 \cite{Durant:2011je,2015MNRAS.452.3357G}.
Unfortunately this source spins with a frequency that is too low that
r-modes could be present. Other spectral analyses \cite{Bogdanov:2006ap,Forestell:2014lza,Prinz:2015jkd}
so far only resolve the hot spot and the hard power law component,
but fail to clearly detect a thermal surface emission. However, even
in the absence of an observed thermal surface component such fits
strongly limit the size of the undetected uniform thermal emission 
and one can give upper bounds on the surface temperature that
have to hold in order to be consistent with the present data. Moreover,
even for sources for which only the total X-ray luminosity is available,
the luminosity likewise limits the temperature of a potential thermal
surface component. In addition to the compiled data in the
literature we also add a more stringent bound for another source via
a detailed X-ray analysis of the nearby pulsar PSR~J1231$-$1411. 

In compact stars the damping and the heating due to the saturation
of unstable global oscillation modes depends strongly on the internal
composition. The same holds for the cooling of the star since certain
forms of matter can feature fast cooling mechanisms. In the absence
of fast cooling photon emission from the surface dominates at the
low temperatures present in millisecond pulsars. However, when fast
neutrino cooling, e.g in exotic forms of matter, is present the bulk
emission dominates, which introduces additional uncertainties related
to the heat transport between the core and the surface. We analyze
both standard and exotic compositions and set robust bounds on the
r-mode amplitude in the considered sources by taking into account
the various uncertainties in the micro- and macro-physics. 

\section{X-ray data}

With the launch of the Fermi satellite there has been a significant
increase in the number of millisecond pulsars that are detected in
gamma-rays, e.g. \cite{2014ARA&A..52..211C}. 
Despite this, the limited spatial resolution of the Large Area Telescope
(LAT) \cite{2009ApJ...697.1071A} combined with the low flux of these objects
limits even the identification of the X-ray counterparts in the first
place, let alone constraining the surface emission. Therefore, apart from a few nearby exceptions (e.g. PSR J0437$-$4715) X-ray data of MSPs suffer from low count statistics, which shows itself in the uncertainties
of the temperature and/or flux values reported. Still, the continuous discoveries of MPSs led to searches of
counterparts at other wavelengths and therefore a
wealth of archival data, especially in the X-rays, became available.
Here, we compile the recently published results of these observations
for a number of rapidly rotating, millisecond pulsars. 

In Table \ref{msp_res} we present the limits or measurements of flux $F$,
temperature $T$, distance $D$, luminosity $L_\infty$, spin frequency $f$ and its derivative $\dot f$
for each pulsar, together with appropriate uncertainties, if also reported
by the authors (see the caption for references). These values were obtained with detectors onboard Chandra, XMM$-${\it Newton}, Swift and other X-ray satellites. Most often only the thermal emission from a hot spot on the surface can be detected. This fact shows itself in the inferred apparent radius values, which are also given in the Table \ref{msp_res}. The flux and the corresponding luminosity values presented in Table \ref{msp_res} only refer to the thermal component, if it is detected. Note that in some cases instead of a blackbody temperature, authors report the temperature values inferred from neutron star atmosphere models, specifically the NSA model \cite{1996A&A...315..141Z}. In this model the neutron star is assumed to have a thin atmosphere in hydrostatic and radiative equilibrium and composed only of completely ionized hydrogen. With such a model the photon energy dependence of the opacity can be calculated and the effective temperature of the surface emission can be inferred instead of a color temperature. Neutron star atmosphere models generally result in a broader, harder X-ray spectrum than a pure blackbody emission (see e.g. \cite{2013RPPh...76a6901O} for a recent review) since higher energy photons from deeper layers can also be observed from the atmosphere. Therefore fitting the same dataset with such an atmosphere model results in a lower temperature and larger apparent radius compared to just fitting with a blackbody model. Furthermore assuming a mass and radius these models also take into account the gravitational redshift of the neutron star. In Table \ref{msp_res} we marked such measurements.

\begin{figure}
\includegraphics[scale=0.68]{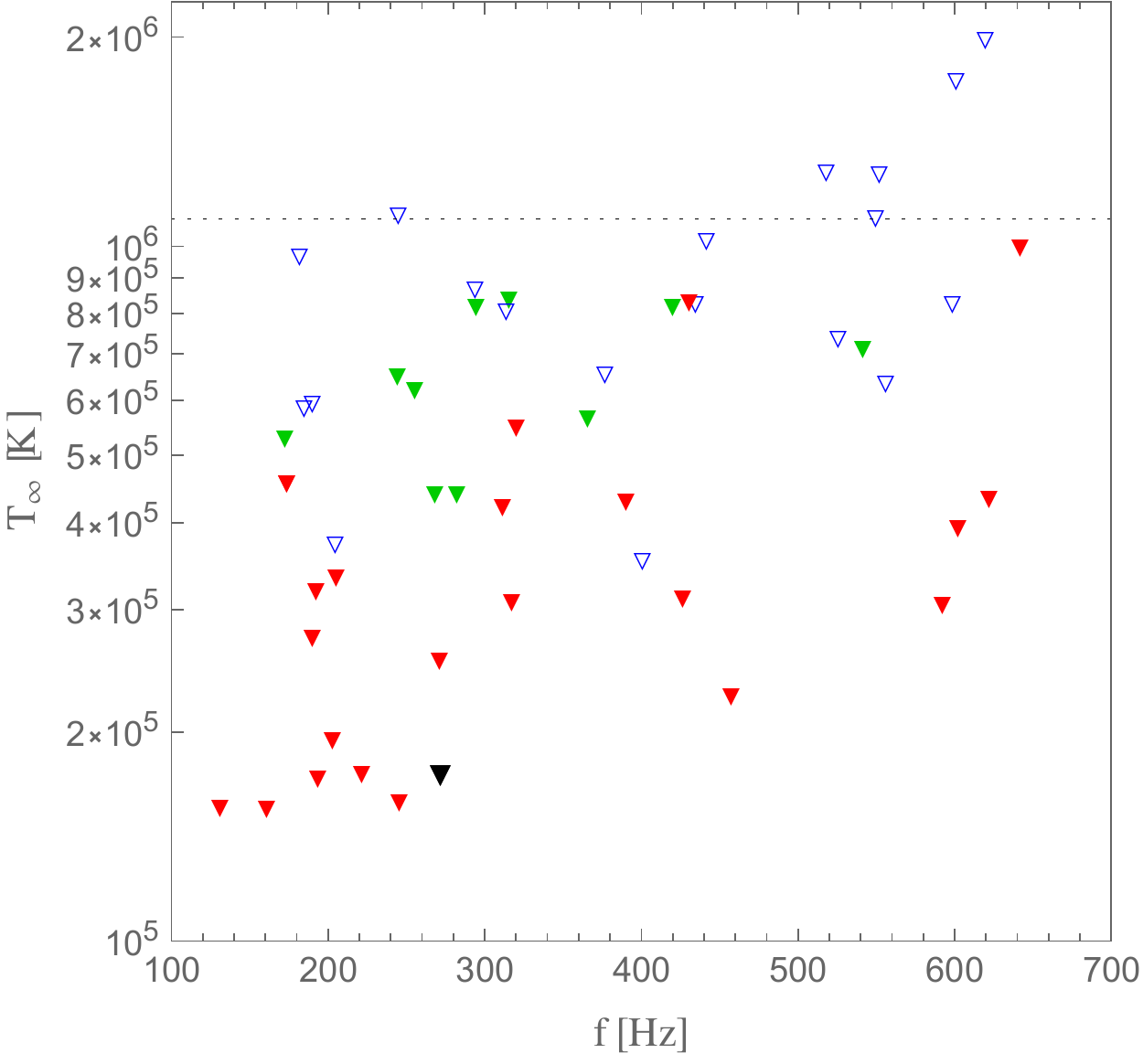}
\caption{\label{fig:temperature-bounds}Bounds on the surface temperatures red-shifted to infinity
for different classes of sources: solid triangles show bounds for
recycled millisecond pulsars and open (blue) triangles bounds for
LMXBs (Mahmoodifar \& Strohmeyer(2013)). The millisecond pulsar bounds stem
both from bounds in the literature on the total luminosity (red), that are compiled in table~\ref{msp_res}, as well as from the recent survey (Prinz \& Becker(2015)) (green). The larger (black) triangle shows the improved
bound for PSR J1231$-$1411 discussed in section \ref{sec:analysis}. The
dotted horizontal line shows for a neutron star with standard modified
Urca cooling ($1.4\,{\rm M_{\odot}}$ model with an APR equation of
state (Akmal et al.(1998))  given in (Alford et al.(2010))) the temperature
below which surface photon cooling dominates over bulk neutrino emission.}
\end{figure}

In addition the complementary data compiled in the recent survey by Prinz and Becker
\cite{Prinz:2015jkd} is shown in tab.~\ref{tab:temperature-sources}.
These tables are supplemented by timing data from the ATNF \cite{Manchester:2004bp}.
The corresponding temperature bounds are compared to bounds for LMXBs
\cite{Haskell:2012} in Fig.~\ref{fig:temperature-bounds}. As can
be seen the temperature bounds obtained from millisecond pulsars (red
and green full triangles) are systematically lower than the bounds and
measurements from LMXBs (blue open triangles), which are heated by accretion. In addition there seems to be a mild correlation, that faster spinning sources could be slightly warmer which might point to a spin-dependent heating mechanism in these very old sources.

We would like to note and expand the discussion on two interesting sources, the first one is \pulsarsec, which is a transitional binary. Bogdanov et al. showed that the observed X-ray spectra favor a model which includes a thermal and a non-thermal component \cite{2011ApJ...742...97B}. The thermal component is fit by the authors with the NSA model and is attributed to the surface emission of the neutron star. However the derived best fit parameters of the NSA model shows a significant variation in the orbital phase resolved spectroscopy, which is attributed to multi-temperature thermal spectrum, arising due to non-uniform heating across the face of the magnetic polar cap \cite{2011ApJ...742...97B}. For a conservative estimate, we nevertheless use here the total (phase averaged) spectral fit since the errors in the phase resolved fits are sizable and even though the colder phase seems to be fit by a thermal spectrum with a typical neutron star radius, interpreting this as surface emission is problematic since the luminosity from this large emitting area would be considerably larger than the luminosity in the phase where the fit yields a small hot spot and which would be expected to dominate. 
Also, in Section \ref{sec:analysis} we present a detailed analysis of the archival X-ray observations of \pulsar to both present it as an example of how the spectral modeling is performed and also to demonstrate a statistical method to put even stronger limits on the surface emission of MSPs using actual data instead of results from the literature.

\begin{table*}
\centering \caption{Observational and derived properties of the millisecond pulsars ($f>$~100~Hz)
as compiled from literature.}
\begin{tabular}{cccccccccccc}
\hline 
Source   & Type$^{1}$  & $f$ & $-\dot{f}$ $^{2}$ & $\cal F$ $^{3}$ & $M$ & $T$ & $R$ & Flux$^{4}$  & Distance  & $L_\infty$  & Ref. \tabularnewline
 &  & (Hz) &  & ($10^{-4}$) &($M_{\odot}$) & (keV)  & (km)  & ($10^{-14}\frac{\rm erg}{\rm s \, cm}$) & (kpc)  & ($10^{31} \frac{\rm erg}{\rm s}$) & \tabularnewline \tabularnewline
\hline 
\hline 
J1628$-$3205  & R  & $311.5$ & \textendash{} & \textendash{} & \textendash{}  & \textendash{}  & \textendash{}  & \textendash{}  & 1.2  & $<$2.2  & 1\tabularnewline
J1810$+$1744  & BW  & $602.4$ & \textendash{} & \textendash{} &\textendash{}  & \textendash{}  & \textendash{}  & 2.75$\pm$0.71  & 2.0  & 1.32$\pm$0.34  & 2\tabularnewline

J2129$-$0429  & R  & $131.2$ & \textendash{} & \textendash{} & \textendash{}  & 0.16$\pm$0.07  & 0.09  & 0.42  & 0.9  & 0.041  & 3 \tabularnewline
B1937$+$21  & BW  & $641.9$ & $429.1$ & 87 &\textendash{}  & 0.9$\pm$0.1  & \textendash{} & 23.0$\pm$3.0  & 5.0  & 68.8  & 4 \tabularnewline
J0218$+$4232  & BW  & $430.5$ & $143.4$ & 185 &\textendash{}  & \textendash{}  & \textendash{}  & \textendash{}  & 2.7  & 33.0  & 4 \tabularnewline
B1957$+$20  & BW  & $622.1$ & $65.22$ & 21 &\textendash{}  & 0.18$\pm$0.03  & 0.1  & 2.0  & 2.5$\pm$1.0  & 1.5$\pm$0.96  & 5 \tabularnewline
J1731$-$1847  & BW  & $426.5$ & $46.22$ & 11 & \textendash{}  & \textendash{}  & \textendash{}  & 0.83$\pm$0.41  & 2.5  & 0.622$\pm$0.31  & 2 \tabularnewline
J1311$-$3430  & BW  & $390.6$ & $31.98$ & 65 &\textendash{}  & \textendash{} & \textendash{}  &  8.6$\pm$1.4  & 1.4  & 2.02$\pm$0.33  & 2 \tabularnewline
B1257$+$12  & \textendash{}  & $160.8$ & $29.57$ & 2.9 &\textendash{}  & 0.22  & 0.8  & \textendash{}  & 0.5  & 0.03$\pm$0.01  & 6 \tabularnewline
J2214$+$3000  & BW  & $320.6$ & $14.4$ & 473 &\textendash{}  & \textendash{}  & \textendash{}  & 15.85$\pm$7.59  & 1.5  & 4.27$\pm$2.04  & 2\tabularnewline
J1024$-$0719  & \textendash{}  & $193.7$ & $6.95$ & 15 &\textendash{}  & 0.155$\pm$0.034$^{5}$  & 0.1  & \textendash{}  & 0.39  & 0.04$\pm$0.02  & 7 \tabularnewline
J2051$-$0827  & BW  & $221.8$ & $6.266$ & 16 &\textendash{}  & 0.25$\pm$0.05  & $\sim$0.1  & 0.49  & 1.04  & 0.0634  & 8 \tabularnewline
J1012$+$5307  & \textendash{}  & $190.27$ & $6.201$ & 115 &$1.8\pm0.1$ & \textendash{}  & \textendash{}  & 12.00$\pm$0.0  & \textendash{}  & 0.39  & 9 \tabularnewline
J1744$-$1134  & BW  & $245.4$ & $5.382$ & 12 &\textendash{}  & 0.272$\pm$0.098  & 0.03  & 2.72$\pm$0.2  & 0.357  & 0.041$\pm$0.003  & 10 \tabularnewline
J2302$+$4442  & \textendash{}  & $192.59$ & $4.933$ & 263 &\textendash{}  & 0.069$\pm$0.014$^{5}$  & 2.7  & 3.10$\pm$0.40  & 1.18$\pm$0.2  & 0.516$\pm$0.206  & 11 \tabularnewline
J1614$-$2230  & -- & $317.4$ & $9.692$ & 70 & $1.93\pm0.02$ & 0.15$\pm$0.04  & 0.55  & 2.29  & 1.3$\pm$0.25  & 0.463$\pm$0.161  & 12 \tabularnewline
J2124$-$3358  & -- & $202.8$ & $8.46$ & 20 &\textendash{}  & 0.112$\pm$0.009$^{5}$  & 0.32  & \textendash{}  & 0.27  & 0.1$\pm$0.02  & 7 \tabularnewline
J2241$-$5236  & -- & $457.3$ & $13.88$ & 9.7 &\textendash{}  & \textendash{}  & \textendash{}  & 5.22$\pm$0.72  & 0.5  & 0.156$\pm$0.022  & 13 
 \tabularnewline
J1023$+$0038  & TB & 592.4 & 42 & 8.3 & 1.7$\pm$0.2 & 0.065$\pm0.022^{6}$  & $2.5^{+8.2}_{-1.1}$ & -- & 1.3 & 0.6 & 14
 \tabularnewline
J0437$-$4715 & WDB & 173.7 & 17.28 & 1.6 & 1.76$\pm$0.20 & 0.032$_{-0.005}^{+0.007}$ & 11$_{-5}^{+23}$ & -- & 0.156$\pm$0.001 & -- & 15
\tabularnewline
J0030$+$0451 & -- & 205.5 & 4.30 & 338 & -- & 0.034$_{-0.006}^{+0.052}$ & 4.3$_{-2.1}^{+2.8}$ & 40$\pm$0.4 & 0.3 & 0.431$\pm$0.431 & 16
\tabularnewline
\hline 
\label{msp_res}  &  &  &  &  &  &  &  &  &  & \tabularnewline
\end{tabular}\\
{\footnotesize{}{ }{\footnotesize \par}
\begin{raggedright}
{\footnotesize{}$^{1}$ BW : Black Widow, R : Redback, TB : Transitional Binary, WDB : White Dwarf Binary}\\
{\footnotesize{}$^{2}$ Spindown rate in units of ($10^{-16}\frac{1}{{\rm s}^2}$).
}\\

{\footnotesize{}$^{3}$ Derived spindown fraction due to r-modes, see sec.~\ref{sec:stability} for details.
}\\
{\footnotesize{}$^{4}$ Unabsorbed fluxes are given in units of $10^{-14}$~erg/s/cm$^{2}$
and correspond only to the thermal component whose temperature is
also given in the table, if available. Otherwise it represents the total flux.}\\
{\footnotesize{} 
$^{5}$ Reported temperature values are effective redshifted temperatures measured at infinity using neutron star atmosphere models. \\
}
$^{6}$ Reported temperature values are effective unredshifted temperatures measured
using neutron star atmosphere models. \\
\par\end{raggedright}{\footnotesize \par}
{\footnotesize{}References : (1) \cite{2014ApJ...795...72L}; (2) \cite{2015ApJ...814...90A};
(3) \cite{2015ApJ...801L..27H}; (4) \cite{2014ApJ...787..167N}; (5) \cite{2012ApJ...760...92H};
(6) \cite{2007ApJ...664.1072P}; (7) \cite{2006ApJ...638..951Z}; (8) \cite{2012ApJ...748..141W};
(9) \cite{2004A&A...419..269W}; (10) \cite{2011ApJ...733...82M}; (11) \cite{2011ApJ...732...47C}; (12) \cite{2012A&A...544A.108P}; (13)
\cite{2012arXiv1205.1748M}; (14) \cite{2011ApJ...742...97B}; (15) \cite{2015MNRAS.452.3357G}; (16) \cite{2009ApJ...703.1557B}.}\\
{\footnotesize{}All timing data is taken from the ATNF database \cite{Manchester:2004bp}
and where available masses are taken from \cite{Antoniadis:2016hxz}.} }{\footnotesize \par}
\end{table*}

\begin{table}
\begin{tabular}{ccccc}
\hline 
Source & $f$ & $-\dot{f}$ & $M$ & $T_{\infty}^{\left({\rm bound}\right)}$\tabularnewline
 & (Hz) & ($10^{-16}\,{\rm s}{}^{-2}$) & $M_{\odot}$ & ($10^{5}\,{\rm K}$)\tabularnewline
\hline 
\hline 
J0337$+$1715 & $365.953$ & $23.66$ & $1.438\pm0.001$ & $5.6$\tabularnewline
J1103$-$5403 & $294.75$ & $3.2$ & \textendash{}  & $8.1$\tabularnewline
J1719$-$1438 & $172.707$ & $2.399$ & \textendash{}  & $5.2$\tabularnewline
J1751$-$2857 & $255.436$ & $7.347$ & \textendash{}  & $6.2$\tabularnewline
J1843$-$1113 & $541.81$ & $28.15$ & \textendash{}  & $7.1$\tabularnewline
J1853$+$1303 & $244.391$ & $5.206$ & \textendash{}  & $6.5$\tabularnewline
J1933$-$6211 & $282.212$ & $2.947$ & \textendash{}  & $4.4$\tabularnewline
J1946$+$3417 & $315.444$ & $3.682$ & $1.83\pm0.01$ & $8.3$\tabularnewline
J2043$+$1711 & $420.189$ & $9.252$ & \textendash{}  & $8.1$\tabularnewline
J2129$-$5721 & $268.359$ & $15.02$ & \textendash{}  & $4.4$\tabularnewline
\hline 
\end{tabular}
\caption{\label{tab:temperature-sources}Millisecond sources ($f\!>\!100\,{\rm Hz}$)
from the X-ray survey (Prinz \& Becker(2015)) supplemented
by timing data from the ATNF database (Manchester et al.(2004)) and
where available by masses from (Antoniadis et al.(2016))}
\end{table}

\section{X-ray analysis of PSR~J1231-1411}
\label{sec:analysis}
A typical X-ray analysis of a pulsar involves fitting the observed X-ray spectra with a thermal and non-thermal component and as can be seen in Table \ref{msp_res} the thermal component often refers to the hot spot on the surface in the case of MSPs. However in the search for the surface temperature of the neutron star a more careful analysis can be performed, which at least in some cases can yield stronger constraints. Here, we apply such a technique as a proof of concept to one MSP, namely  \pulsar. In a future work we aim to present a systematic approach to the whole population of MSPs for which X-ray data are available.

\pulsar~was discovered by Fermi LAT \cite{2011ApJ...727L..16R} as one of brightest gamma-ray MSPs in the sky ($F_{100}= 10.57_{-0.39}^{+0.62} \times 10^{-8}~{\rm {ph}~{\rm {cm}^{-2}~{\rm {s}^{-1}}}}$), and has a spin period of 3~ms~(271.4~Hz). The observed period
derivative implies a surface magnetic field strength of ($2-3)\times10^{8}$~G and a spin-down luminosity of $\sim2\times10^{34}\,{\rm erg \, s}^{-1}$. \pulsar~is in a binary system, which has an orbital period of 1.9 days and the mass of the companion is $\sim0.2-0.3M_{\odot}$, consistent with the orbital period white dwarf mass relation of \cite{1995MNRAS.273..731R}. \cite{2011ApJ...727L..16R} also report the results of a XMM-\textit{Newton} observation of this source. The X-ray spectrum can be modeled using a blackbody with a temperature of 0.21~keV and the corresponding unabsorbed X-ray flux of this blackbody is $F_{bb}=15(_{-7.4}^{+5.3})\times10^{-14}{\rm {erg}\,{\rm {cm}^{-2}\,{\rm {s}^{-1}}}}$ in the 0.5$-$8~keV range. Assuming a distance of 0.4~kpc, this flux corresponds to an X-ray luminosity of ${L}_{\rm bb}=2.9(_{-1.4}^{+1.0})10^{30}$~erg~s$^{-1}$. The observed spectrum can also be fitted with a power-law model with a photon index of $\Gamma=4.23(_{-0.38}^{+0.41})$ and column density of ${\rm {nH}=1.8(_{-0.5}^{+0.6})\times10^{21}}\,{\rm cm}^{-2}$.

\pulsar~has been observed several times with SUZAKU, CHANDRA, XMM-\textit{Newton}, and Swift satellites. We re-analyzed all of the archival observations of this source, using the Chandra, SUZAKU and XMM-\textit{Newton} data to obtain a constraint on the surface emission. A log of these observations is given in Table \ref{obs_table}.

For the calibration of the XMM-\textit{Newton} data we used SAS v14.0 together with the available calibration files as of September 2015. The source region is extracted from a circle with a radius of 640 pixels and the background region is selected from the same CCD with a similar size. EPIC-pn, MOS1 and MOS2 spectra were grouped to have at least 25 counts per spectral channel and not to oversample the energy resolution of the instrument by at least a factor of 3. Chandra data was calibrated and the X-ray spectrum is extracted using CIAO 4.6 and the \textit{specextract} tool within this package. The source and background regions are selected from circles with a radius of $\sim4^{\prime\prime}$. The spectrum is grouped to have at least 15 counts per spectral channel. There were two SUZAKU observations in the archives, for this study we used both of these observations and the data from XIS0 and XIS3. The SUZAKU XIS data was calibrated using the \textit{aepipeline} tool distributed with the HEASOFT v6.17 and the CALDB database\footnote{http://heasarc.gsfc.nasa.gov/FTP/caldb/}. The X-ray spectra are extracted using Xselect, which uses the \textit{xisrmfgen} and \textit{xissimarfgen}  tools to generate the necessary response files. XIS data were also grouped to have at least 25 counts per spectral channel using the \textit{grppha} tool.

\begin{figure}
\centering \includegraphics[scale=0.4]{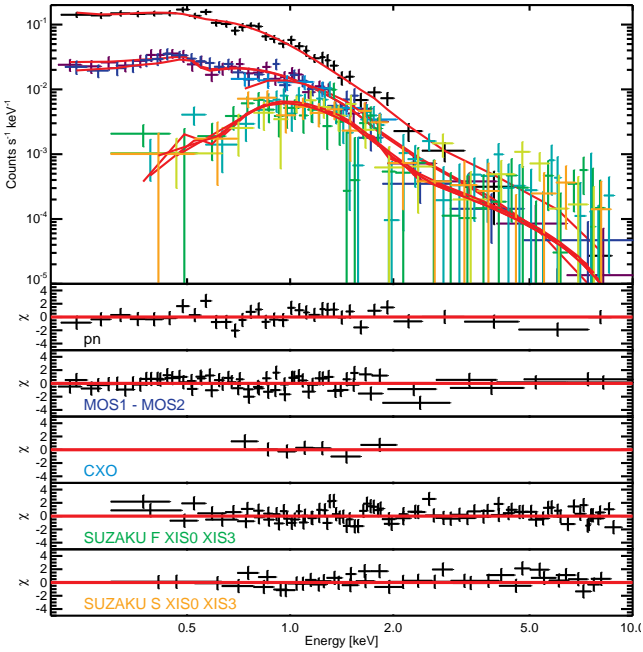} -\caption{Observed X-ray count spectra of \pulsar~obtained by various satellites and detectors. Different responses of the instruments result in a different observed spectral shape. The lower panels show the residuals from the best fit model, which is
a blackbody plus a power-law model, for each satellite, detector, (pn, MOS1, and MOS2 refer to different CCDs onboard XMM-Newton. CXO refers to the Chandra X-ray Observatory ACIS-I observation)  or
in the case of SUZAKU data, individual observations (F and S refer to the
first and second SUZAKU observations). Colors of the labels in lower panels denote the corresponding data in the upper panel.}
\label{spec_res} 
\end{figure}

We fit all the spectra simultaneously using Xspec version 12.9.0 \cite{1996ASPC..101...17A}. In a simultaneous fit the assumed model is convolved with the response of each dataset and the corresponding responses, where the model parameters are linked between the different datasets. So as a result it is possible to get one global $\chi^2$ value and a set of best fit parameters that represent all of the observed data, taking into account the individual responses. The data can be easily fit with a blackbody plus a power-law model ($\chi^{2}$ = 0.94556 for 224 degrees of freedom) in the 0.2$-$10.0~keV range with all the model parameters linked. To take into account the effect of the interstellar absorption, we used the \textit{tbabs} model \cite{2000ApJ...542..914W} assuming interstellar abundances and fixed its value to the weighted average of the hydrogen column density as given by \cite{2005A&A...440..775K} in the direction of the source (nH=3.45$\times10^{20}{\rm {cm}^{-2}}$). Note that when allowed to be free, this parameter can not be constrained with the existing data. Similar to the early results, the resulting parameters show a relatively hot surface emission from a very small region on the surface. The best fit blackbody temperature and emitting area are found as ${\rm {kT}=0.158_{-0.009}^{+0.008}}$~keV, ${\rm {R}=0.144_{-0.021}^{+0.024}}$~km, respectively. The uncertainties represent the 1-$\sigma$ uncertainties of the measurements and a distance of 0.4~kpc is assumed to calculate the radius. The X-ray spectra, best fit model and its residuals for each instrument are shown in Figure \ref{spec_res}.

\begin{figure}
\centering \includegraphics[scale=0.4]{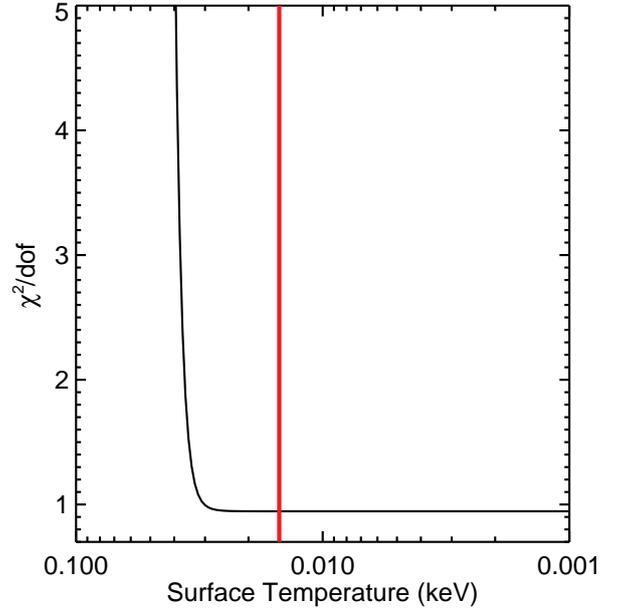} -\caption{Variation of the temperature of the cool blackbody component as a
function of $\chi^{2}/dof$ assuming that this component is coming from the whole surface of the neutron star. The red vertical line shows
the temperature where $\chi^{2}$ starts to become larger than the best fit value found using a blackbody plus a power-law model with the existing data. At lower temperatures the $\chi^{2}/dof$ is constant since the additional blackbody component does not generate detectable flux and is a fixed component, hence does not change the degrees of freedom of the fit.}
\label{kt_res} 
\end{figure}

The inferred blackbody temperature and the apparent emitting area
indicates a hot spot on the surface of the neutron star with a temperature
of 1.8$\times10^{6}$ K. In order to put a stronger limit on the cooling of
this neutron star, we tried to add another blackbody component with
an emitting area of 10~km (assuming the same distance). Any such
addition with one or more free parameters increases the $\chi^{2}$.
If we only allow the temperature to vary and fit the spectrum with
this one additional free parameter, the best fit $\chi^{2}$ becomes
0.94980 for 223 degrees of freedom and the temperature of the cool component can be found as 21675~K. Note that the resulting $\chi^{2}$ value is still worse than the earlier two component model. So this additional
blackbody component has no effect on the fit to the data apart from
changing the degrees of freedom, showing that we can not directly measure the surface temperature of the neutron star within the limits of the current data set. Clearly with this data only an upper-limit can be put on the surface emission of the temperature. For this purpose, we fixed the temperature of the cool blackbody component too and set it to different values. Specifically, we created a thousand step loop and set the temperature to values between 0.001$-$1.0~keV for a fixed radius of 10~km and fit the spectra with this additional fixed component, which does not change the degrees of freedom of the fit. Obviously a temperature value that is high enough to create a significant flux, causes the resulting $\chi^{2}$ to be worse than the initial blackbody
plus a power-law model, since such a component is statistically not needed within
the current data but if the flux of this component is low enough the $\chi^{2}$ / dof would return to its initial value since the parameters of this component are fixed and do not change the degrees of freedom. Practically we determined the highest temperature blackbody with the assumed normalization (radius and distance) that would result in a flux that would not affect the calculated $\chi^{2}$. Performing such an analysis yields an upper limit of the temperature as kT=0.015~keV. The resulting variation in the $\chi^{2}/dof$ as a function of blackbody temperature is shown in Figure \ref{kt_res}. Note that this is the inferred blackbody temperature, where the gravitational redshift of
the neutron star and any possible effects of an atmosphere are not taken into account. Also the distance is kept fixed at 0.4~kpc.

\begin{table*}
\centering \caption{Log of observations of J1231$-$1411.}
\begin{tabular}{cccccc}
\hline 
Satellite  & Detector  & Date  & OBSID  & Mode  & Exposure \tabularnewline
 &  &  &  & (s)  & \tabularnewline
\hline 
\hline 
XMM-\textit{Newton}  & EPIC pn  & 2009-07-15  & 0605470201 & Extended Full Frame  & 18950\tabularnewline
XMM-\textit{Newton}  & EPIC MOS1  & 2009-07-15  & 0605470201 & Full Frame  & 29040\tabularnewline
XMM-\textit{Newton}  & EPIC MOS1  & 2009-07-15  & 0605470201  & Full Frame & 29080\tabularnewline
Chandra  & ACIS-I  & 2012-12-14  & 15362  & VFAINT  & 9936 \tabularnewline
SUZAKU  & XIS0, XIS3 & 2009-07-28  & 804017020  & \textendash{}  & 52640 \tabularnewline
SUZAKU  & XIS0, XIS3 & 2009-07-08  & 804017010  & \textendash{}  & 26310 \tabularnewline
\hline 
\label{obs_table} &  &  &  &  & \tabularnewline
\end{tabular}
\end{table*}

\section{r-mode amplitude bounds}

r-modes \cite{Papaloizou:1978zz,Andersson:1997xt,Friedman:1997uh,Lindblom:1998wf,Andersson:1998ze,Andersson:2000mf} are global toroidal oscillations of rotating stars that are driven unstable by gravitational wave emission via the Friedman-Schutz mechanism \cite{Friedman:1978hf}. The magnitude of the r-mode amplitude determines their impact on the evolution of a compact star. There are different definitions of a dimensionless r-mode amplitude in the literature
and we use the definition $\alpha$ given in \cite{Lindblom:1998wf}.

Whereas initially very large amplitudes were assumed \cite{1998PhRvD..58h4020O}, stringent bounds
on the r-mode amplitude have by now been obtained from different data
sets. For millisecond pulsars their timing data imposes a direct limit
on the r-mode amplitude since these sources would have to spin down
faster than observed if the r-mode amplitude were too large. Due to
the extreme precision of the timing data, this spindown limit \cite{2010PhRvD..82j4002O}
yields the most precise predictions, but unfortunately turned out
to be the least restrictive of the different limits proposed. Nevertheless,
the most constraining source puts a limit of $\alpha\lesssim10^{-7}$
on the dimensionless r-mode amplitude.

Thermal data sets analogous bounds on the amplitude, since due to
the significant dissipation to saturate the mode sources would be
hotter than observed if the r-mode amplitude were too high. So far
thermal X-ray data for LMXBs has been analyzed and more stringent
bounds on the r-mode amplitude of a few times $10^{-8}$ were obtained
when assuming fiducial values for the unknown source parameters \cite{Mahmoodifar:2013quw,Alford:2013pma}.
 Here we derive analogous bounds from millisecond pulsars. Such thermal
bounds involve significant uncertainties from bulk source parameters,
spectral modeling, thermal transport, ... , and we will estimate these
uncertainties to obtain robust upper bounds.

The r-mode physics is described by the energy transfer between the
different components of the system, namely the rotational energy,
the mode energy, the thermal energy and the energy emitted into gravitational
waves. The power pumped into the r-mode, that is for a saturated mode
eventually dissipated into heat, is given by

\begin{equation}
P_{G}\equiv\frac{dE_{{\rm mode}}}{dt}=\hat{G}\left(1+3\frac{1-\chi}{\chi}\right)\chi^{6}\alpha^{2}\Omega^{8}\label{eq:r-mode-power}
\end{equation}
where the constant $\hat{G}$ is determined by static source properties

\begin{equation}
\hat{G}\equiv\frac{2^{17}\pi}{3^{8}5^{2}}\tilde{J}^{2}GM^{2}R^{6}\label{eq:G-hat}
\end{equation}
The weakly frequency dependent factor $\chi\!\left(\Omega\right)\approx1$
describes the deviation of the connection between the rotation frequency
$f\!=\!\Omega/\left(2\pi\right)$ and the r-mode oscillation frequency
$\nu\!=\!\omega/\left(2\pi\right)$ from their canonical relation
$\omega\!=\,-\frac{4}{3}\chi\!(\Omega)\Omega$ and is determined by
general relativistic and rotation corrections \cite{Kokkotas:2015gea}. The power dissipated
in the star as well as its cooling luminosity likewise have, at least
over certain parameter regimes, general power law forms

\begin{equation}
P_{D}=\hat{D}T^{\delta}\lambda^{\Delta}\Omega^{\psi}\quad,\quad L=\hat{L}T^{\theta}\lambda^{\Theta}\label{eq:power-laws}
\end{equation}
As shown in \cite{Alford:2012yn}, the fast thermal and slow spin-down
evolution in the presence of r-modes happen on very different time
scales. Therefore, these sources are in a thermal steady state where
r-mode heating balances the radiative cooling $P_{G}=L$. Taking into
account that there could be other heating mechanisms in a compact
star, a temperature measurement or upper limit
imposes an upper bound on the r-mode saturation amplitude

\begin{equation}
\alpha_{{\rm sat}}\leq\sqrt{\frac{L}{\left(3-2\chi\right)\chi^{5}\hat{G}\Omega^{8}}}~\label{eq:general-saturation-amplitude-bound}
\end{equation}
where the thermal luminosity $L=\sum_{i}L_{i}$ can have contributions from different cooling processes. For standard neutron stars without enhanced cooling mechanisms, like direct Urca
neutrino emission, photon cooling from the surface dominates over
modified Urca neutrino emission for temperatures roughly below $10^{6}\,{\rm K}$.
Taking into account that the red-shifted luminosity at infinity is related to the corresponding temperature via $L_\infty \!=\! \pi^3/15 R^{2} T_\infty^{4}$, in this case the above bound reduces to 

\begin{equation}
\alpha_{{\rm sat}}\leq\sqrt{\frac{3^{7}5}{2^{25}\pi^{6}\left(3-2\chi\right)\chi^{5}G}}\frac{1}{\tilde{J}MR^{2}}\left(1-\frac{2GM}{R}\right)^{-1}\frac{T_{\infty}^{2}}{f^{4}}\label{eq:saturation-amplitude-bound}
\end{equation}
I.e. in addition to the important temperature dependence of these
bounds, fast spinning sources also lead to significantly lower
limits. If fast cooling processes are absent, eq.~(\ref{eq:saturation-amplitude-bound})
is a good approximation for most millisecond pulsars since the power
law exponents for photon ($\theta_{\gamma}=4$) and neutrino emission
($\theta_{\nu}>4$) are rather different, so that for modified Urca
processes ($\theta=8$) the neutrino emission is suppressed by a factor
16 compared to photon emission already at about $5\times10^{5}\,{\rm K}$,
i.e. at half the temperature where they are of equal size. Yet, if
fast cooling processes would be present they could generally compete with
photon cooling in observed sources.

To obtain a rigorous upper amplitude bound, we note that the dominant
general relativistic corrections to the factor $\chi>1$ increase
it so that a conservative upper bound is obtained when setting $\chi=1$.
Similarly a robust upper bound is obtained for minimal values of $\tilde{J}$,
$M$ and $R$. The factor $\tilde{J}$, which characterizes the compactness
of the source, has been shown to be bounded by $\tilde{J}>1/\left(20\pi\right)$
\cite{Alford:2012yn}, the mass of a compact star is generally larger
than a solar mass $M>M_{\odot}$ \cite{Antoniadis:2016hxz} and the radius larger than about $R>10\,{\rm km}$ \cite{2014ApJ...784..123L,2016ApJ...820...28O}. 
This yields the rigorous upper bound on the saturation
amplitude in a given source in the absence of fast cooling

\begin{equation}
\alpha_{{\rm sat}}<1.40\times10^{-9}\left(\frac{T_{\infty}}{10^{5}\,{\rm K}}\right)^{2}\left(\frac{500\,{\rm Hz}}{f}\right)^{4}\label{eq:numeric-saturation-amplitude-bound}
\end{equation}
Note that although we used extreme values to obtain robust bounds,
eq.~(\ref{eq:numeric-saturation-amplitude-bound}) imposes nevertheless
very low amplitudes compared to the bounds stemming from the spin-down
limit which yields for the most restrictive source $\alpha_{{\rm sat}}\lesssim10^{-7}$.
The bound eq.~(\ref{eq:numeric-saturation-amplitude-bound}) holds
both for colder LMXBs \cite{Mahmoodifar:2013quw,Alford:2013pma} and non-accreting
millisecond pulsars. It is clear that millisecond pulsars can set
lower bounds on the r-mode amplitude since they are not heated by
accretion and correspondingly are expected to be colder.

\begin{figure}
\includegraphics[scale=0.68]{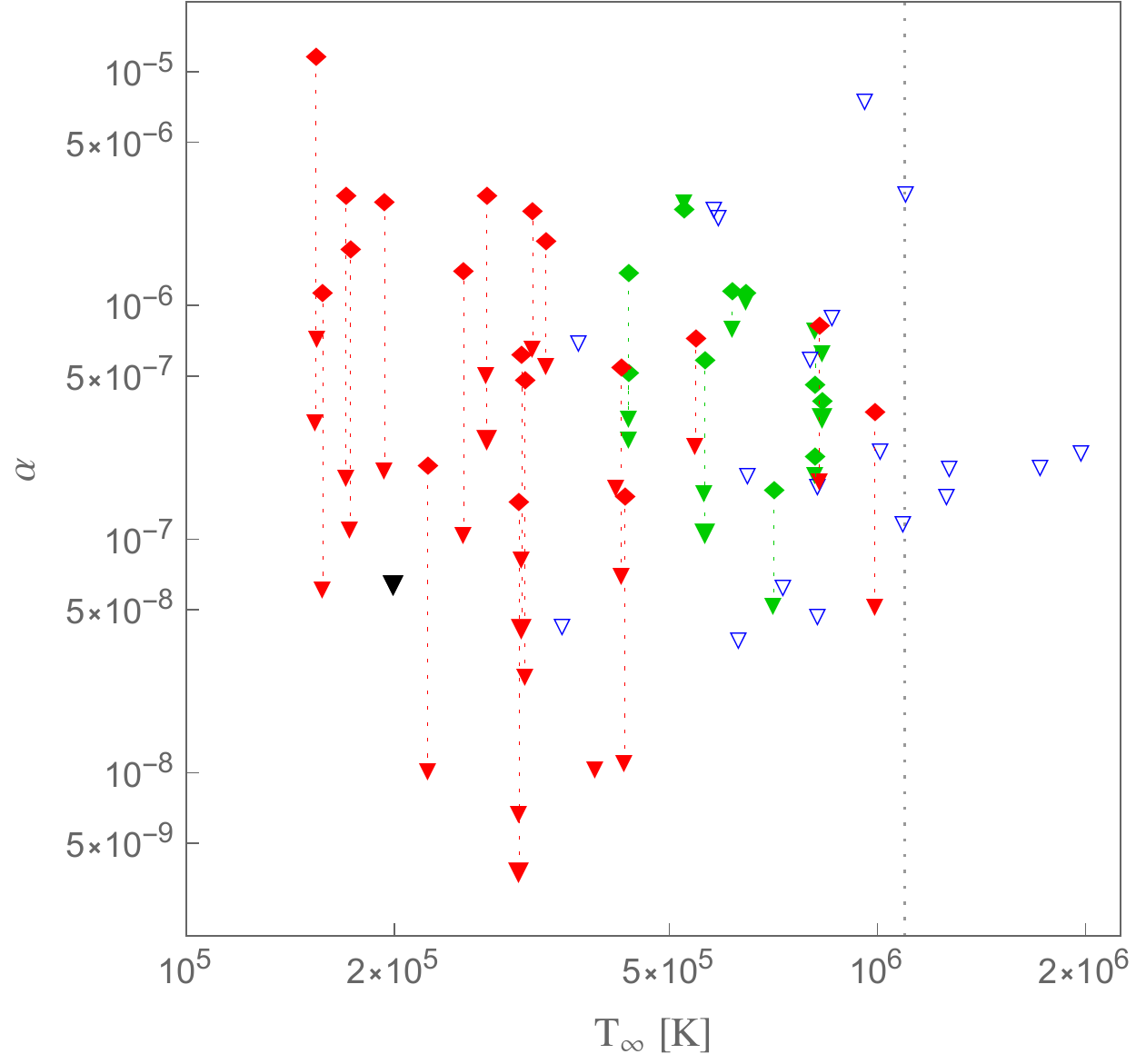}\caption{\label{fig:amplitude-bounds}Bounds on the r-mode amplitude stemming from thermal X-ray data (triangles) compared to those from pulsar
timing data (diamonds, where available) for different sets of sources. Temperature
bounds are obtained from the absence of a thermal surface component
for sources where at least a thermal hot spot and/or a power law component
could be spectrally fitted (black) and the total count rate / luminosity
for sources where the data was not good enough to perform a spectral
analysis (red and green solid symbols), whereas the corresponding
larger symbols show the enhancement due to the measurement of the
source mass. The vertical dotted line segments are included to guide the eye which
data points belong to the same source. 
In addition data for accreting LMXBs is shown (blue open
triangles). The dotted vertical line shows again the temperature below
which surface photon cooling dominates.}
\end{figure}

The bounds for the different sources with thermal X-ray data are shown
in fig.~\ref{fig:amplitude-bounds} (solid triangles). For a few sources actual mass measurements are available \cite{Antoniadis:2016hxz}
which enhance the above bounds and the corresponding data points are
shown by the larger triangles. As can be
seen the luminosity bounds from some millisecond pulsars set very
tight bounds on the r-mode amplitude\textemdash the most restrictive
bound $\alpha_{{\rm sat}}\lesssim 4\times10^{-9}$ being obtained
for the transitional binary \pulsarsec, where the particularly low bound is partly achieved due to the mass measurement for this source. Even though we performed a conservative estimate, this source, as well as the black widow pulsar PSR B1957$+$20, is slightly more uncertain and therefore it is ensuring that other sources are just at or even slightly below $10^{-8}$, like J2241$-$5236 and J1810$+$1744. In case of the latter two sources the total luminosity is used so that the bound on the surface temperature is unambiguous. In particular J2241$-$5236 is also very close, so that the distance measurement should be more precise and interstellar absorption is negligible, eliminating the remaining uncertainties we could not  systematically take into account here.

These values are substantially lower than the best bound stemming from LMXBs (open triangles) \cite{Mahmoodifar:2013quw,Alford:2013pma}. As noted before to make them robust the bounds we give here are very conservative. To reach their high frequencies, the millisecond sources we study need to have been spun up by accretion in a binary. The mass transfer should therefore result in star masses in the upper range of possible mass values, as is clearly seen for those sources in tab.~\ref{msp_res} where mass measurements are available. Contrasting this to our very conservative assumption $M>M_{\odot}$, likely values should be significantly lower. Taking further into account that the dimensionless parameter $\tilde J$ is likewise for heavier stars significantly above its absolute lower limit further strengthens the amplitude bounds. For instance for a standard $2\ M_\odot$ neutron star model \cite{Alford:2010fd} with an APR equation of state \cite{Akmal:1998cf}, as had been used in \cite{Mahmoodifar:2013quw}, we find that the r-mode amplitude bounds are more than a factor of two smaller, the lowest one setting the limit $\alpha_{{\rm sat}}\lesssim 10^{-9}$.

The obtained amplitudes are not too
far from the regime $\alpha_{{\rm sat}}\lesssim10^{-10}$ where r-modes could be completely
ruled out in theses sources, since even the fastest millisecond pulsars
could cool out of the instability region \cite{Alford:2013pma}. What
is more is that these amplitudes are many orders magnitude below those
that well established saturation mechanisms, like for instance mode-coupling
\cite{Arras:2002dw,Bondarescu:2013xwa}, can provide. Therefore, neutron
stars with minimal damping require another strong non-linear dissipation
mechanism that can completely damp or saturate r-modes at such low
amplitudes. Such an additional mechanism is not established, and in
the absence of it the minimal picture of neutron stars is at this point
incompatible with the astrophysical data. This points to new physics and there are several interesting proposals \cite{Madsen:1999ci,2009MNRAS.397.1464H,Alford:2013pma,Gusakov:2013jwa,Haskell:2013hja,Alford:2014jha}.

The above analysis was performed for the case that photon emission
from the atmosphere dominates the cooling of the star, which is the
case at the low temperatures present in millisecond pulsars when only
standard neutrino cooling mechanisms, like modified Urca emission,
are present. However, there is the possibility that fast cooling mechanisms,
like pair breaking or even direct Urca emission \cite{2001PhR...354....1Y}, are present in these sources, and in this case neutrino emission would dominate even at
the low temperatures of old millisecond pulsars. 
A stronger cooling mechanism requires for a source with an observed temperature in thermal steady state that there is more heating, which in turn requires a larger r-mode amplitude.
Different cooling mechanisms have a different temperature
power law behavior which drastically changes the form of the thermal
steady state curve in $T$-$\Omega$-space \cite{Mahmoodifar:2013quw,Alford:2013pma}
and therefore this requires a dedicated analysis. In order to gauge
the impact of fast cooling mechanisms we consider the extreme case
of direct Urca emission in hadronic matter which is the fastest known
cooling mechanisms and therefore sets the limit on the impact of fast
cooling. For direct Urca cooling the bound becomes when using the conventions given in \cite{Alford:2010fd}
\[
\alpha_{{\rm sat}}\leq\sqrt{\frac{3^{8}5^{2}\tilde{L}\Lambda_{{\rm QCD}}}{2^{15}\tilde{J}^{2}\Lambda_{{\rm EW}}^{4}GM^{2}R^{3}}}\frac{T^{4}}{\Omega^{4}}
\]
where $T$ is the core temperature of the star which has to be connected
to the observed temperature via a thermal transport model \cite{1982ApJ...259L..19G} for the
stars envelope and the dimensionless quantity $\tilde{L}$ characterizes
the neutrino emission in the star.

\begin{figure}
\includegraphics[scale=0.68]{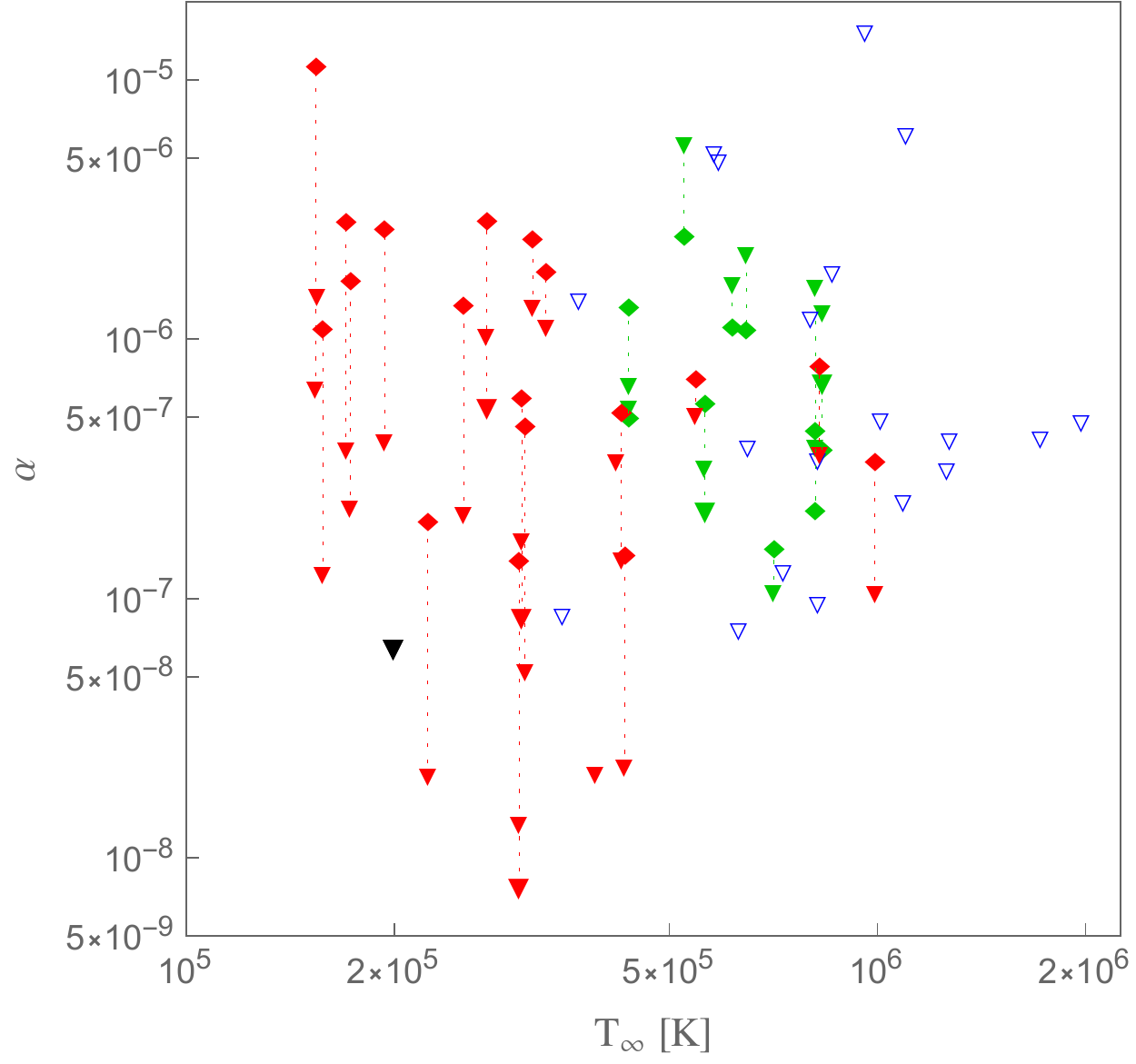}
\caption{\label{fig:amplitude-bounds-fast-ccoling}Bounds on the r-mode amplitude
when direct Urca cooling (Yakovlev et al.(2001)), representing the fastest known cooling mechanism,
is present in the source. The data are the same as in fig.~\ref{fig:amplitude-bounds}, and in this case photon cooling would become relevant only below the shown temperature range.}
\end{figure}

The bounds for the case of direct Urca cooling are shown in fig.~\ref{fig:amplitude-bounds-fast-ccoling} and as can be seen these are very similar to the case of photon cooling.
As before these bounds include the uncertainties arising from the
bulk star properties ($M$, $R$, $\tilde{J}$) but do not include
the additional uncertainties on the microscopic material properties,
like the neutrino emissivity in $\tilde{L}$ and the thermal transport
in the crust \cite{1982ApJ...259L..19G}. Due to our incomplete understanding of
the underlying interactions and of the many-body effects it is harder to estimate these uncertainties, but due to the weak dependence on the corresponding parameters \cite{Alford:2012yn}, a robust bound should not be much weaker.

\section{r-mode stability and spindown fraction}
\label{sec:stability}

\begin{figure}
\includegraphics[scale=0.68]{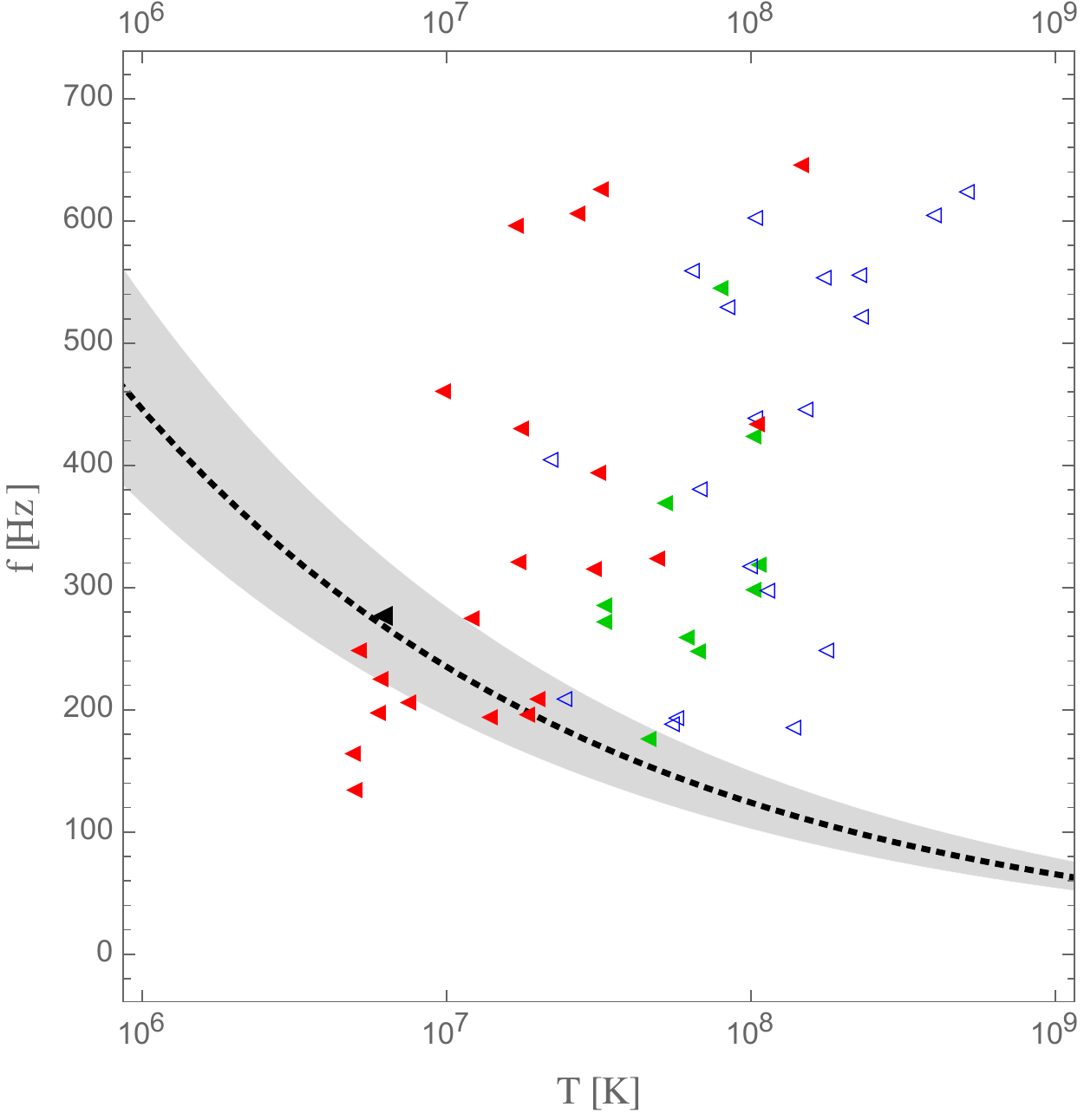}
\caption{\label{fig:instability-bounds}Instability region for hadronic nuclear matter compared to core temperature estimates for the different sources. The gray band shows the uncertainty range on the instability region due to uncertainties in the micro- and macro-physics, cf. (Alford \& Schwenzer(2012) \& (2014).}
\end{figure}

For some sources one can even go a step further and use the bounds on the surface temperature to show that r-modes are completely absent. This is done by comparing the data to the instability region in $T$-$f$-space. Its boundary is determined by the condition $P_G=P_D$ and r-modes are unstable in sources above this curve. Here $T$ is the core temperature of the star which is determined as above from a model for its envelope. Our results are given in fig.~\ref{fig:instability-bounds} where the dashed curve shows the instability region of hadronic nuclear matter in the low temperature regime where dissipation due to shear viscosity dominates \cite{Alford:2010fd} with an error band reflecting the uncertainties \cite{2014NuPhA.931..740A}. As can be seen there are several sources with frequencies up to about 250 Hz that are outside of the r-mode instability region and which correspondingly do not emit gravitational waves due to r-modes. In particular the sources J2129$-$04 and B1257$+$12, for which we give new temperature bounds, are far away from the boundary. The new temperature bounds do, however, not qualitatively change the situation for all the fast spinning sources, which are far inside the instability region and even if we so far only have upper bounds for their temperature they could not have escaped the instability region unless the r-mode saturation amplitude is extremely low $\alpha_{\rm sat} \lesssim 10^{-10}$ \cite{Alford:2013pma}.

For those sources that can be in the instability region we can limit the contribution of r-modes to the observed spindown rate. The angular momentum loss due to r-mode gravitational wave emission is given by
$dJ_{g}/dt=-3\hat{G}\Omega^{7}\alpha^{2}$. This can be compared to the observed angular momentum loss
$dJ_{r}/dt=I\dot{\Omega}$, given in terms of the moment of inertia $I=\tilde{I}MR^{2}$, where the arising dimensionless constant is bounded by $\tilde{I} \geq 2/9$ \cite{Alford:2012yn}. Using the amplitude bound eq.~(\ref{eq:general-saturation-amplitude-bound}), the spindown fraction due to r-modes $\cal F$ is for observed sources bounded by
\begin{equation}
{\cal F}\equiv\frac{dJ_{g}}{dt}/\frac{dJ_{r}}{dt} \leq \frac{3L}{I\Omega\left|\dot{\Omega}\right|} \leq \frac{9\pi T_\infty^{4}}{40 M f\left|\dot{f}\right|} \left(1-\frac{2GM}{R}\right)^{-2}
\end{equation}
where the last relation holds when photon cooling dominates the thermal luminosity $L$. 

The bounds on the spindown fraction $\cal F$ are likewise given in table~\ref{msp_res}. As can be seen in all considered sources r-modes are responsible for less than a few percent of the observed spindown rate, as has been found previously for sources in LMXBs in \cite{Mahmoodifar:2013quw}. Yet, for some sources the stringent bounds on the temperatures in millisecond pulsars set even bounds on the spindown fraction as low as $10^{-4}$. Therefore it is fair to conclude that r-modes are irrelevant for the current spindown evolution of these sources and other mechanisms must be responsible for the observed spindown rate. This demonstrates the generic property of r-modes that even when their amplitude is so low that they are irrelevant for the spindown, they can nevertheless have a significant impact on the thermal evolution---taking into account that the bounds are saturated precisely when the r-modes are the only heating source.

\section{Gravitational wave emission}

Due to their enormous stability, millisecond radio pulsars would be
perfect sources for gravitational wave astronomy \cite{2010PhRvD..82j4002O,Aasi:2013sia,Alford:2014pxa,Kokkotas:2015gea}.
They would allow us to perform searches over long time intervals and
could therefore yield high precision measurements. In the absence of a
direct gravitational wave detection, we have so far only upper bounds
on the gravitational wave emission, stemming from their spin-down data.
These spin-down limits have recently been revised \cite{Alford:2014pxa}
taking into account that the same r-mode saturation mechanism should
be present in all of these sources. This resulted in universal spin-down
limits that could be orders of magnitude smaller. Nevertheless, the
bounds for several sources were close to the sensitivity threshold
for realistic searches with advanced LIGO. Here we will use the thermal
X-ray bounds for millisecond pulsar temperatures and the corresponding
r-mode amplitudes to obtain more stringent limits on the gravitational
wave signal for sources with thermal data.

\begin{figure}
\includegraphics[scale=0.68]{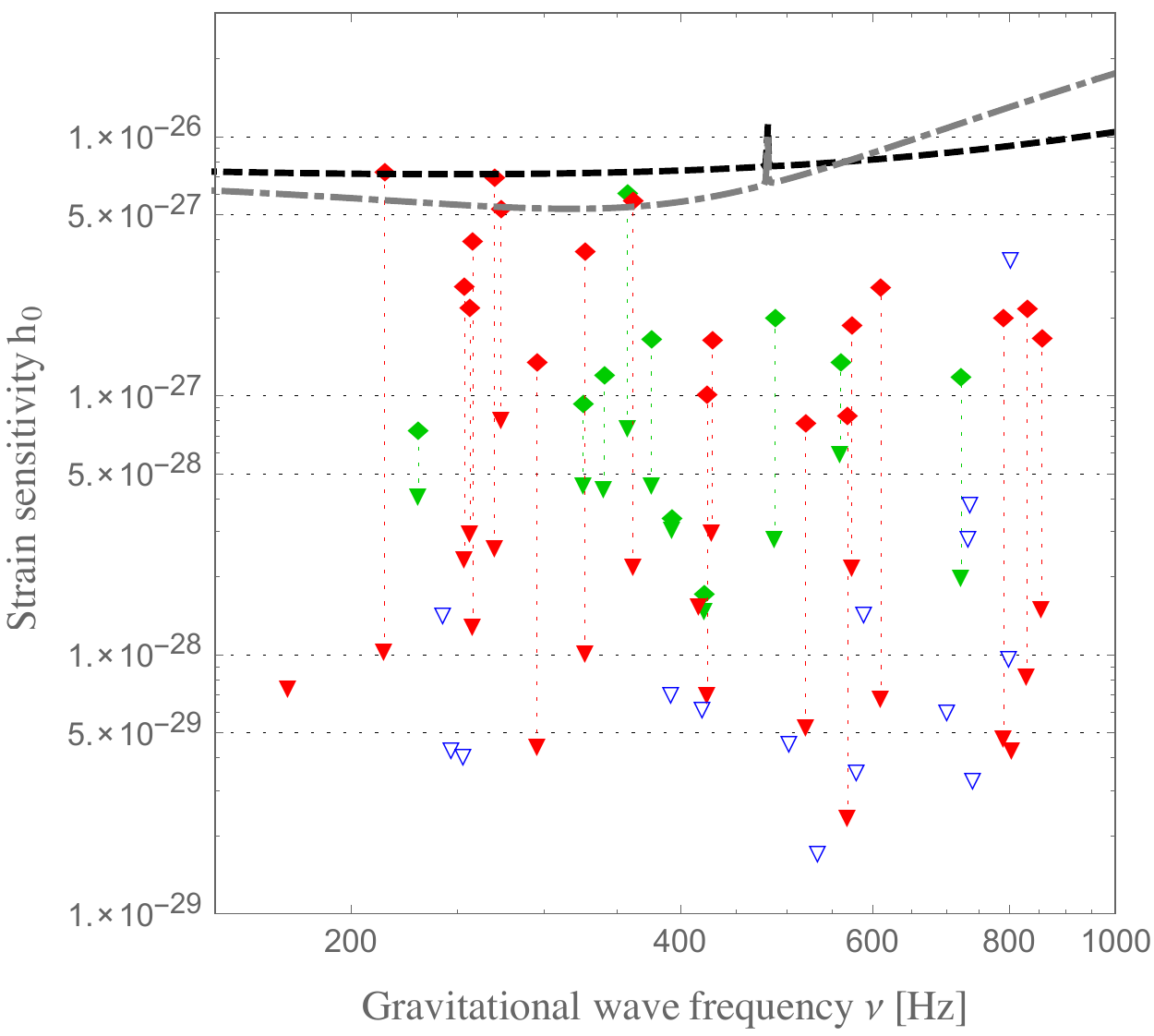}\caption{\label{fig:strain-bounds}Upper bounds on the intrinsic strain amplitude
stemming from thermal X-ray data (triangles) compared to the spin-down
limit (diamonds) from pulsar timing data for different sets of sources.
The vertical dotted line segments are included to guide the eye which
data points belong to the same source. Solid triangles (red and green)
show millisecond sources with temperature bounds whereas open triangles
(blue) show LMXB sources. The larger (black) triangle shows the improved
bound for PSR J1231$-$1411. The results are compared to sensitivity
curves for advanced LIGO for a coherent search using one year of data in both the standard (black, dashed) and the neutron star enhanced configuration (gray, dot-dashed).}
\end{figure}

The gravitational wave signal in a terrestrial detector is characterized
by the intrinsic strain amplitude which is for r-mode emission given
by 

\begin{equation}
h_{0}=\sqrt{\frac{2^{15}\pi}{3^{6}5}}G\tilde{J}MR^{3}\frac{\chi^{2}\Omega^{3}\alpha}{D}\label{eq:strain-amplitude}
\end{equation}
Using the bounds for the r-mode amplitude above yields upper bounds
on the intrinsic strain amplitude when fast cooling is absent

\begin{equation}
h_{0}\leq\sqrt{\frac{3\pi G}{\left(3-2\chi\right)\chi}}\left(1-\frac{2GM}{R}\right)^{-1}\frac{R}{D}\frac{T_{\infty}^{2}}{f}\label{eq:strain-amplitude-bound}
\end{equation}
The results for the sources discussed in this work are compared to
the sensitivity of advanced LIGO in fig.~\ref{fig:strain-bounds}.
The bounds on the gravitational wave strain stemming from the temperature
bounds in fig.~\ref{fig:temperature-bounds} (solid triangles) are
compared to the spin-down limits stemming from the pulsar timing data
(diamonds). In addition we also show the corresponding thermal bounds
from LMXBs (open triangles) \cite{Mahmoodifar:2013quw,Kokkotas:2015gea}.
As can be seen the thermal limits are below the spin-down limits for
all sources and strengthen the latter by up to two orders of magnitude.
This means that many of these sources are far below the sensitivity
of current detectors. Nevertheless, there are several sources that
are not too far below the detection sensitivity of advanced LIGO and
could be in reach with the combination of the advanced LIGO detectors
with other second generation detectors like advanced Virgo and KAGRA,
or the advent of third generation detectors like the Einstein telescope.

\section{Conclusion}

Using recent X-ray data we set stringent bounds $\alpha \lesssim 10^{-8}$ on the r-mode amplitude in millisecond radio and high energy pulsars. These bounds are considerably
tighter than the spin-down limits obtained from the pulsar timing data
and even lower than those obtained previously for accreting sources
in LMXBs. Correspondingly they show that there must be significant
damping in these sources. However, non of the standard damping mechanism that are known to be present in standard neutron stars can provide such strong damping. An example for such a standard saturation mechanism, that is present independent of the composition, is the coupling of different oscillations modes \cite{Arras:2002dw,Bondarescu:2013xwa}, which is insufficient since it can only saturate r-modes at $\alpha_{\rm sat} \sim 10^{-6}-10^{-5}$. The same holds for Ekman-layer damping, which cannot damp r-modes in fast sources \cite{Alford:2013pma}, as well as for magnetic effects due to differential rotation \cite{Rezzolla:1999he,Friedman:2015iqa}, which likewise have no significant impact 
taking into account the small magnetic fields present in millisecond pulsars \cite{Rezzolla:pc}.

Therefore, the new bounds strongly increase the discrepancy between standard,
well-constrained damping mechanisms and the astrophysical data, so that in absence of such trivial explanations astroseismology becomes an efficient tool to study the internal composition of compact stars. Structurally more complicated stars, involving for instance multi-component superfluids \cite{2009MNRAS.397.1464H,Haskell:2013hja}, could have the potential to explain the data, while even more exotic compositions like ungapped quark matter \cite{Madsen:1999ci,Alford:2013pma} or hybrid stars \cite{Alford:2014jha} have already been shown to be fully consistent with the present data. 

Our analysis also sets limits as low as $\lesssim 10^{-4}$ on the spindown fraction due to r-modes that underline the generic feature, that even when r-modes are so small that they are irrelevant for the spindown of a source, they nevertheless can have a significant impact on its thermal evolution. The obtained bounds also impose analogous bounds on the gravitational wave strain in current interferometers, like
advanced LIGO, which rule out the possibility to detect the continuous
gravitational wave emission from these sources without further enhancements
or third generation detectors.

We also demonstrated a simple yet effective method to put an observational limit on the surface emission of these objects, using \pulsar~as an example case. In a future study we will expand this search by taking into account the measurement uncertainties in the distances, radii and the amount of interstellar extinction for a number of pulsars for which there is high quality data. We note that the NICER mission that will be launched to the International Space Station in early 2017 (see e.g. \cite{2014SPIE.9144E..20A}), may have a significant contribution to the search of surface emission of these objects thanks to its large effective area in the soft X-ray band.

\section*{Achnowledgements}
We are grateful to Sebastien Guillot for useful discussions on the X-ray spectral analysis of PSR J0437-4715.

\bibliographystyle{mn2e}

\begin{thebibliography}{}

\bibitem[Aasi et al.(2014)]{Aasi:2013sia} Aasi, J., Abadie, J., Abbott, B.~P., et al.\ 2014, \apj, 785, 119

\bibitem[Akmal et al.(1998)]{Akmal:1998cf} Akmal, A., Pandharipande, V.~R., \& Ravenhall, D.~G.\ 1998, \prc, 58, 1804

\bibitem[Alford et al.(2012)]{Alford:2010fd} Alford, M.~G., Mahmoodifar, S., \& Schwenzer, K.\ 2012, \prd, 85, 024007 

\bibitem[Alford \& Schwenzer(2014)]{Alford:2012yn} Alford, M.~G., \& Schwenzer, K.\ 2014, \apj, 781, 26 

\bibitem[Alford \& Schwenzer(2014)]{Alford:2013pma} Alford, M.~G., \& Schwenzer, K.\ 2014, Physical Review Letters, 113, 251102

\bibitem[Alford \& Schwenzer(2014)]{2014NuPhA.931..740A} Alford, M.~G., \& Schwenzer, K.\ 2014, Nuclear Physics A, 931, 740

\bibitem[Alford et al.(2015)]{Alford:2014jha} Alford, M.~G., Han, S., \& Schwenzer, K.\ 2015, \prc, 91, 055804 

\bibitem[Alford \& Schwenzer(2015)]{Alford:2014pxa} Alford, M.~G., \& Schwenzer, K.\ 2015, \mnras, 446, 3631

\bibitem[Andersson(1998)]{Andersson:1997xt} Andersson, N.\ 1998, \apj, 502, 708

\bibitem[Andersson \& Kokkotas(2001)]{Andersson:2000mf} Andersson, N., \& Kokkotas, K.~D.\ 2001, International Journal of Modern Physics D, 10, 381

\bibitem[Andersson et al.(1999)]{Andersson:1998ze} Andersson, N., Kokkotas, K., \& Schutz, B.~F.\ 1999, \apj, 510, 846 

\bibitem[Antoniadis et al.(2016)]{Antoniadis:2016hxz} Antoniadis, J., Tauris, T.~M., Ozel, F., et al.\ 2016, arXiv:1605.01665 

\bibitem[Arnaud(1996)]{1996ASPC..101...17A} Arnaud, K.~A.\ 1996, Astronomical Data Analysis Software and Systems V, 101, 17 

\bibitem[Arras et al.(2003)]{Arras:2002dw} Arras, P., Flanagan, E.~E., Morsink, S.~M., et al.\ 2003, \apj, 591, 1129

\bibitem[Arumugasamy et al.(2015)]{2015ApJ...814...90A} Arumugasamy, P., Pavlov, G.~G., \& Garmire, G.~P.\ 2015, \apj, 814, 90 

\bibitem[Arzoumanian et al.(2014)]{2014SPIE.9144E..20A} Arzoumanian Z., et al., 2014, SPIE, 9144, 914420 

\bibitem[Atwood et al.(2009)]{2009ApJ...697.1071A} Atwood, W.~B., Abdo, A.~A., Ackermann, M., et al.\ 2009, \apj, 697, 1071 

\bibitem[Becker \& Tr{\"u}mper(1999)]{Becker:1998yf} Becker, W., \& Tr{\"u}mper, J.\ 1999, \aap, 341, 803 

\bibitem[Bogdanov et al.(2006)]{Bogdanov:2006ap} Bogdanov, S., Grindlay, J.~E., Heinke, C.~O., et al.\ 2006, \apj, 646, 1104 

\bibitem[Bogdanov \& Grindlay(2009)]{2009ApJ...703.1557B} Bogdanov, S., \& Grindlay, J.~E.\ 2009, \apj, 703, 1557 

\bibitem[Bogdanov et al.(2011)]{2011ApJ...742...97B} Bogdanov, S., Archibald, A.~M., Hessels, J.~W.~T., et al.\ 2011, \apj, 742, 97 

\bibitem[Bondarescu \& Wasserman(2013)]{Bondarescu:2013xwa} Bondarescu, R., \& Wasserman, I.\ 2013, \apj, 778, 9

\bibitem[Caraveo(2014)]{2014ARA&A..52..211C} Caraveo, P.~A.\ 2014, araa, 52, 211 

\bibitem[Cognard et al.(2011)]{2011ApJ...732...47C} Cognard, I., Guillemot, L., Johnson, T.~J., et al.\ 2011, \apj, 732, 47 

\bibitem[Durant et al.(2012)]{Durant:2011je} Durant, M., Kargaltsev, O., Pavlov, G.~G., et al.\ 2012, \apj, 746, 6 

\bibitem[Forestell et al.(2014)]{Forestell:2014lza} Forestell, L.~M., Heinke, C.~O., Cohn, H.~N., et al.\ 2014, \mnras, 441, 757

\bibitem[Friedman et al.(2016)]{Friedman:2015iqa} Friedman, J.~L., Lindblom, L., \& Lockitch, K.~H.\ 2016, \prd, 93, 024023

\bibitem[Friedman \& Morsink(1998)]{Friedman:1997uh} Friedman, J.~L., \& Morsink, S.~M.\ 1998, \apj, 502, 714 

\bibitem[Friedman \& Schutz(1978)]{Friedman:1978hf} Friedman, J.~L., \& Schutz, B.~F.\ 1978, \apj, 222, 281

\bibitem[Gudmundsson et al.(1982)]{1982ApJ...259L..19G} Gudmundsson, E.~H., Pethick, C.~J., \& Epstein, R.~I.\ 1982, \apjl, 259, L19 

\bibitem[Guillot et al.(2015)]{2015MNRAS.452.3357G} Guillot, S., et. al, 2015, arXiv:1512.03957 

\bibitem[Gusakov et al.(2014)]{Gusakov:2013jwa} Gusakov, M.~E., Chugunov, A.~I., \& Kantor, E.~M.\ 2014, Physical Review Letters, 112, 151101

\bibitem[Haskell et al.(2009)]{2009MNRAS.397.1464H} Haskell, B., Andersson, N., \& Passamonti, A.\ 2009, \mnras, 397, 1464 

\bibitem[Haskell et al.(2012)]{Haskell:2012} Haskell, B., Degenaar, N., \& Ho, W.~C.~G.\ 2012, \mnras, 424, 93 

\bibitem[Haskell et al.(2014)]{Haskell:2013hja} Haskell, B., Glampedakis, K., \& Andersson, N.\ 2014, \mnras, 441, 1662

\bibitem[Huang et al.(2012)]{2012ApJ...760...92H} Huang, R.~H.~H., Kong, A.~K.~H., Takata, J., et al.\ 2012, \apj, 760, 92 

\bibitem[Hui et al.(2015)]{2015ApJ...801L..27H} Hui, C.~Y., Hu, C.~P., Park, S.~M., et al.\ 2015, apjl, 801, L27 

\bibitem[Kalberla et al.(2005)]{2005A&A...440..775K} Kalberla, P.~M.~W., Burton, W.~B., Hartmann, D., et al.\ 2005, aap, 440, 775 

\bibitem[Kokkotas \& Schwenzer(2015)]{Kokkotas:2015gea}
Kokkotas, K.~D.,  Schwenzer, K.\ 2015, Eur.\ Phys.\ J.\ A, 52, 38

\bibitem[Lattimer \& Steiner(2014)]{2014ApJ...784..123L} Lattimer, J.~M., \& Steiner, A.~W.\ 2014, \apj, 784, 123

\bibitem[Linares(2014)]{2014ApJ...795...72L} Linares, M.\ 2014, \apj, 795, 72 

\bibitem[Lindblom et al.(1998)]{Lindblom:1998wf} Lindblom, L., Owen, B.~J., \& Morsink, S.~M.\ 1998, Physical Review Letters, 80, 4843 

\bibitem[Madsen(2000)]{Madsen:1999ci} Madsen, J.\ 2000, Physical Review Letters, 85, 10 

\bibitem[Mahmoodifar \& Strohmayer(2013)]{Mahmoodifar:2013quw} Mahmoodifar, S., \& Strohmayer, T.\ 2013, \apj, 773, 140 

\bibitem[Manchester et al.(2005)]{Manchester:2004bp} Manchester, R.~N., Hobbs, G.~B., Teoh, A., \& Hobbs, M.\ 2005, \aj, 129, 1993

\bibitem[Marelli et al.(2011)]{2011ApJ...733...82M} Marelli, M., De Luca, A., \& Caraveo, P.~A.\ 2011, \apj, 733, 82 

\bibitem[Marelli(2012)]{2012arXiv1205.1748M} Marelli, M.\ 2012, arXiv:1205.1748 

\bibitem[Ng et al.(2014)]{2014ApJ...787..167N} Ng, C.-Y., Takata, J., Leung, G.~C.~K., Cheng, K.~S., \& Philippopoulos, P.\ 2014, \apj, 787, 167 

\bibitem[Owen et al.(1998)]{1998PhRvD..58h4020O} Owen, B.~J., Lindblom, L., Cutler, C., et al.\ 1998, \prd, 58, 084020 

\bibitem[Owen(2010)]{2010PhRvD..82j4002O} Owen, B.~J.\ 2010, \prd, 82, 104002

\bibitem[{\"O}zel(2013)]{2013RPPh...76a6901O} {\"O}zel, F.\ 2013, Reports on Progress in Physics, 76, 016901 

\bibitem[{\"O}zel et al.(2016)]{2016ApJ...820...28O} {\"O}zel, F., Psaltis, D., G{\"u}ver, T., et al.\ 2016, \apj, 820, 28

\bibitem[Pancrazi et al.(2012)]{2012A&A...544A.108P} Pancrazi, B., Webb, N.~A., Becker, W., et al.\ 2012, aap, 544, A108 

\bibitem[Papaloizou \& Pringle(1978)]{Papaloizou:1978zz} Papaloizou, J., \& Pringle, J.~E.\ 1978, \mnras, 182, 423 

\bibitem[Pavlov et al.(2007)]{2007ApJ...664.1072P} Pavlov, G.~G., Kargaltsev, O., Garmire, G.~P., \& Wolszczan, A.\ 2007, \apj, 664, 1072

\bibitem[Prinz \& Becker(2015)]{Prinz:2015jkd} Prinz, T., \& Becker, W.\ 2015, arXiv:1511.07713 

\bibitem[Ransom et al.(2011)]{2011ApJ...727L..16R} Ransom, S.~M., Ray, P.~S., Camilo, F., et al.\ 2011, \apjl, 727, L16 

\bibitem[Rappaport et al.(1995)]{1995MNRAS.273..731R} Rappaport, S., Podsiadlowski, P., Joss, P.~C., Di Stefano, R., \& Han, Z.\ 1995, \mnras, 273, 731 

\bibitem[Rezzolla et al.(2000)]{Rezzolla:1999he} Rezzolla, L., Lamb, F.~K., \& Shapiro, S.~L.\ 2000, \apjl, 531, L139 

\bibitem[Rezzolla(2016)]{Rezzolla:pc} Rezzolla L., private communication 

\bibitem[Webb et al.(2004)]{2004A&A...419..269W} Webb, N.~A., Olive, J.-F., Barret, D., et al.\ 2004, aap, 419, 269 

\bibitem[Wilms et al.(2000)]{2000ApJ...542..914W} Wilms, J., Allen, A., \& McCray, R.\ 2000, \apj, 542, 914 

\bibitem[Wu et al.(2012)]{2012ApJ...748..141W} Wu, J.~H.~K., Kong, A.~K.~H., Huang, R.~H.~H., et al.\ 2012, \apj, 748, 141

\bibitem[Yakovlev et al.(2001)]{2001PhR...354....1Y} Yakovlev, D.~G., Kaminker, A.~D., Gnedin, O.~Y., \& Haensel, P.\ 2001, \physrep, 354, 1

\bibitem[Zavlin et al.(1996)]{1996A&A...315..141Z} Zavlin, V.~E., Pavlov, G.~G., \& Shibanov, Y.~A.\ 1996, aap, 315, 141 

\bibitem[Zavlin(2006)]{2006ApJ...638..951Z} Zavlin, V.~E.\ 2006, \apj, 638, 951 

\end{thebibliography}

\end{document}